\documentclass[12pt,superscriptaddress]{article}

\usepackage{enumerate}
\usepackage{empheq}
\usepackage{booktabs}
\usepackage[english]{babel}
\usepackage{amsmath,amssymb,amsbsy,amstext, amsthm, simplewick}
\usepackage{hyperref}
\usepackage{graphicx}
\usepackage{amsfonts}
\usepackage{color}
\usepackage{framed}
\usepackage[top=3cm, bottom=3cm, left=3cm, right=3cm]{geometry}
\usepackage{verbatim}
\usepackage{cite}

\usepackage{authblk}

\usepackage{colortbl}
\definecolor{summersky}{cmyk}{0.71,0.33,0,0.5}
\definecolor{flamingo}{cmyk}{0,0.51,0.71,0.5}
\definecolor{rp}{cmyk}{0.2, 1, 0.6, 0}
\definecolor{pacificblue}{cmyk}{0.95,0.3,0, 0.5}
\definecolor{gray60}{cmyk}{0.4,0.4,0,0.8}

\def\simleq{\; \raise0.3ex\hbox{$<$\kern-0.75em
      \raise-1.1ex\hbox{$\sim$}}\; }

\def\simgeq{\; \raise0.3ex\hbox{$>$\kern-0.75em
      \raise-1.1ex\hbox{$\sim$}}\; }

\hypersetup{
    pdftoolbar=true,        
    pdfmenubar=true,        
    pdffitwindow=true,     
    pdfstartview={FitH},    
    pdfnewwindow=true,      
    colorlinks=true,       
    linkcolor=pacificblue,          
    citecolor=flamingo,        
    filecolor=magenta,      
    urlcolor=pacificblue           
}

\newcommand{\be}{\begin{eqnarray} }
\newcommand{\ee}{\end{eqnarray} }
\newcommand{\bs}{\begin{split} }
\newcommand{\es}{\end{split} }

\def\be{\begin{eqnarray}}
\def\ee{\end{eqnarray}}

\def\[{\left [}
\def\]{\right ]}
\def\({\left (}
\def\){\right )}

\def\r2{\sqrt{2}}

\graphicspath{{Figures/}}

\begin{document}

\begin{titlepage}

\bigskip\

\vspace{.5cm}
\begin{center}

{\fontsize{25}{28}\selectfont  \sffamily \bfseries  Vacua Morghulis}


\end{center}

\vspace{1cm}

\begin{center}
{\fontsize{13}{30}\selectfont Ben Freivogel$^\diamondsuit$ and Matthew Kleban$^{\spadesuit}$} \end{center}


\begin{center}

\vskip 8pt
\textsl{$^\diamondsuit$ GRAPPA and ITFA, Institute of Physics, Universiteit van Amsterdam, \\
Science Park 904, 1090 GL Amsterdam, Netherlands}
\vskip 7pt

\textsl{$^\spadesuit$ Center for Cosmology and Particle Physics, Department of Physics, New York University, United States of America}
\vskip 7pt

\end{center}

\vspace{1.2cm}
\hrule \vspace{0.3cm}
\noindent {\sffamily \bfseries Abstract} \\[0.1cm]
We conjecture that all vacua in string theory decay. More precisely, non-supersymmetric vacua in string theory are at most metastable and eventually decay, while supersymmetric vacua are only marginally stable.
 All de Sitter vacua with small cosmological constant are metastable to decay via a process that occurs on scales small compared to the horizon size.
 \vskip 10pt
\hrule

\vspace{0.6cm}
 \end{titlepage}

\section{Introduction}

In \cite{ArkaniHamed:2006dz}, Arkani-Hamed, Motl, Nicolis, and Vafa proposed a bound on the charge/mass ratio of the lightest charged particle: $m \leq e$ (in units where an extremal black hole has $m/e=1$).  The bound was motivated by the requirement that no non-supersymmetric black hole should be absolutely stable.

String theory seems to contain a vast number of perturbatively stable vacua. A given vacuum may be able to decay to a lower-energy vacuum by nucleating a bubble of true vacuum. More generally, vacua can be connected to other vacua by domain walls.   It is natural to propose a universal bound on the tension of  domain walls, motivated by the requirement that no non-supersymmetric vacuum be exactly stable. 

Hence we make the following conjectures about string theory vacua that are well-described by semi-classical Einstein gravity:
\begin{itemize}
\item Every non-supersymmetric Minkowski and anti-de Sitter (AdS) vacuum is at best metastable, meaning it can decay to at least one other lower energy vacuum.  Every supersymmetric vacuum is marginally unstable, meaning it is connected to at least one other vacuum by a domain wall that saturates the stability bound.  
\item All de Sitter  vacua are  at best metastable,  decaying at a rate much faster than the Poincare recurrence rate. Further, when the horizon size is large compared to every other length scale (such as the Kaluza-Klein and supersymmetry breaking scales), the decay occurs on scales small compared to the horizon size. 
\end{itemize}
We present two major lines of argument for these conjectures: by analogy to the weak gravity conjecture, and by imposing a consistency requirement that emerged in studying the dynamics of eternal inflation.

If correct, our conjectures have important consequences for understanding the string theory landscape.  If all Minkowski and AdS vacua are at most marginally stable, solutions exist where an arbitrarily large region of the lower-energy vacuum is contained in a spacetime that is asymptotically in the higher-energy vacuum. These domain walls are flat when the vacuum is supersymmetric and marginally stable; otherwise they are expanding. De Sitter vacua are all below the ``great divide" \cite{Bousso:2006am, Aguirre:2006ap}, meaning that the decay channel persists in the limit that the cosmological constant is taken to zero.  In the context of eternal inflation, this implies that their dynamics is \emph{not} described by de Sitter thermal equilibrium.

\paragraph{Note:} While this work was in progress, a  paper by Ooguri and Vafa appeared \cite{Ooguri:2016pdq} that has substantial overlap with ours.  We present some  arguments complementary to theirs, and make a  more general conjecture that applies to de Sitter vacua (such as the one we presumably find ourselves in) along with both supersymmetric and non-supersymmetric anti-de Sitter. Danielsson, Dibitetto, and Vargas \cite{Danielsson:2016rmq} have argued for  the opposite conclusion---that every perturbatively stable AdS vacuum is non-perturbatively stable. If their conclusions are correct, our conjecture is wrong.

\section{Arguments from dualizing the weak gravity conjecture.}
Our interest in this subject was inspired by \cite{Hebecker:2015zss}. For some potentials, one can dualize the scalar to a flux (for a particularly simple example, the massive sine-Gordon model in 1+1 dimensions is dual to the Schwinger model). Domain walls connecting minima of the scalar potential become charged branes under this duality (kinks of the sine-Gordon model become electrons in the Schwinger model), and the  weak gravity conjecture bounds the tension of such branes. Therefore, at least for these special potentials, the weak gravity conjecture bounds the tension of  domain walls (and thus the decay instanton action) from above (see also \cite{Hebecker:2016dsw}).

This suggests that the weak gravity conjecture could place a bound on instanton actions for generic potentials. The original weak gravity conjecture was motivated by demanding that all black holes can decay. A natural analogue is to demand that all vacua can decay.  In the limit that the spacetime is nearly flat on both sides of the domain wall, and the two vacua are related by a change in flux, this was in fact suggested in the original paper \cite{ArkaniHamed:2006dz}.
Here we are simply extending the conjecture to situations where the cosmological constant changes, potentially by a large amount, across the domain wall. 

\subsection{Anti-BPS logic}
The weak gravity conjecture is a kind of anti-BPS bound: it says the mass of a certain charged particle must be {\it smaller} or equal to the minimum value it could have in a supersymmetric theory. Domain walls between two supersymmetric vacua obey a BPS bound on the tension of the domain wall, 
\be \label{bps}
T \geq 2 \left| \Delta (e^{K/2} W) \right|
\ee
where $T$ is the tension of the domain wall, $W$ is the superpotential,  $K$ is the Kahler potential, and the right side of the equation is the difference between the two vacua.\footnote{We use the conventions of \cite{Cvetic:1996vr}, with $8 \pi G = c = \hbar = 1$.}

On the other hand, the  higher-energy vacuum can decay to the lower via domain wall nucleation if
\be \label{stab}
T < {2 \over \sqrt{3}} \left(  \sqrt{|V_1|}  -  \sqrt{|V_2|}  \right)
\ee
where $V_{i}$ is the value of the potential at the minimum corresponding to the vacuum and we define $|V_{1}| > |V_{2}|$ (in the supersymmetric case, $V_i= -3 e^{K_i} |W_i|^2$) \cite{Coleman:1980aw}.

The BPS bound \eqref{bps} coincides with the stability bound \eqref{stab}  if the two vacua have the same phase of the superpotential $W$ \cite{Cvetic:1992st}. This is the case in simple situations, such as for the $D3$ brane domain walls connecting two different $AdS_5 \times S_5$ vacua with different values of the flux. In this case, BPS domain walls saturate the stability bound, and the upper vacuum is marginally stable.  

When a vacuum is marginally stable in this sense, solutions exist where a {\it flat} domain wall interpolates between the true vacuum and the false vacuum. As we demonstrate in Sec.~\ref{lands}, this implies that an arbitrarily large region of the true vacuum can be constructed within a region of false vacuum at the cost of an energy per unit area of the domain wall that scales to zero in the large volume limit.

BPS domain walls can  exist between two vacua that do not have the same phase of the superpotential, but their intrinsic geometry generically is not flat. A BPS wall has a tension $T$  that is bounded by the triangle inequality \cite{Cvetic:1996vr}:
\be
{2 \over \sqrt{3}} \left( \sqrt{|V_1|} - \sqrt{|V_2|} \right)  \leq T = 2 \left|  \Delta\ (e^{K/2} W) \right| \leq  {2 \over \sqrt{3}} \left( \sqrt{|V_1|}  +  \sqrt{|V_2|} \right).
\ee
The minimum value of the tension corresponds to the flat domain walls discussed above: a solution which approaches the true vacuum in the infrared (away from the boundary) and the false vacuum in the UV (near the boundary). The domain walls that take the maximum tension allowed for a BPS wall are also flat \cite{Cvetic:1992st}. However, these geometries have no UV region: they consist of two IR regions, glued across a high-tension domain wall. 
In between these two extreme cases, the intrinsic geometry of the domain wall is that of a lower-dimensional AdS spacetime. 

If we write AdS in terms of lower-dimensional AdS slices, the metric takes the form
\be
ds^2 = d\rho^2 +\ell^2 \cosh^2(\rho/\ell) dA^2
\ee
$\ell$ is the AdS radius, and $dA^2$ is the metric on a Lorentzian-signature AdS spacetime of the appropriate dimensionality (our discussion here is in general dimension).  To construct the solutions, in the thin-wall approximation, two different AdS spacetimes with different values of the AdS radius $\ell$ should be glued along a surface of $\rho={\rm constant}$, with the constant determined by the Israel junction conditions.  

Beyond the thin wall approximation, the metric takes the form
\be
ds^2 = d\rho^2 + g^2(\rho) dA^2
\ee
with the metric function $g(\rho)$ approaching the form for the two different AdS spacetimes asymptotically, 
\be
g(\rho) \to 
\begin{cases}
\ell_1 \cosh(\rho/\ell_1)\ {\rm for}\ \rho \to \infty\\
\ell_2 \cosh(\rho/\ell_2)\ {\rm for} \ \rho \to -\infty
\end{cases}
\ee

The main point here is that BPS domain walls do not necessarily saturate the stability bound \eqref{stab}---and for  walls that do not saturate \eqref{stab}, small perturbations that break supersymmetry will not  destabilize the false vacuum.  Therefore, at least for supersymmetric vacua, it is not the case that the tension for the domain wall between \emph{any} two vacua is bounded above by the instability bound \eqref{stab}, and it is not reasonable that this would be true for non-supersymmetric vacua either.  Indeed, this would be somewhat analogous to the very strong requirement that \emph{every} particle in the spectrum satisfies the weak gravity conjecture $ g > m$.\footnote{In the context of the weak gravity conjecture for particles, Heidenreich, Reece, and Rudelius have argued for a strengthened form of the conjecture \cite{Heidenreich:2016aqi}.  It would be interesting to investigate whether such a strengthened form exists for vacuum decays as well.}

  Instead, we propose the weaker condition that every  vacuum is at best marginally stable.
More precisely, every non-supersymmetric vacuum is connected to at least one lower energy vacuum by a domain wall with tension satisfying \eqref{stab}:
\be  \label{conj}
0 \leq T <   {2 \over \sqrt{3}} \left( \sqrt{| V_1 | } -  \sqrt{| V_2 | } \right),
\ee
while every supersymmetric vacuum is connected to at least one other vacuum by a BPS domain wall saturating \eqref{conj}.
This is somewhat reminiscent of the ``weak weak gravity conjecture''---that there must exist \emph{some} particle (not necessarily the lightest) with $g > m$.

It is worth noting that the bound derived in \cite{Hebecker:2015zss} for inflationary axion models
$$
T \simleq e  \sim m f 
$$ 
agrees  with \eqref{conj} up to $\mathcal{O}(1)$ factors.  The axion potential considered there is $V = {1 \over 2} m^{2} \phi^{2} + \alpha\cos{\phi / f}$, so the potential energy in each minimum is $V_n \approx{1 \over 2} (m 2 \pi n f )^{2}$.  Hence,  $\sqrt{V_{n}} - \sqrt{V_{n-1}} \approx {2 \pi \over \sqrt{2}} m f$, in agreement with \eqref{conj}.\footnote{M.K. thanks F.~Rompineve for discussions on this point.}

The bound \eqref{conj} is well-defined in the thin-wall limit. Beyond that limit, it is defined by demanding that an arbitrarily large region of the true vacuum can be embedded in the false vacuum with finite energy.  Additionally, in the case of de Sitter, our conjecture implies that the decay rate is less than the Poincare recurrence time, and that the length scales associated with the decay are less than the de Sitter Hubble length.  In the language of \cite{Aguirre:2006ap}, all de Sitter vacua are below the ``great divide."

\paragraph{AdS as near-horizon limit.}
Many AdS spacetimes can be realized as the near-horizon limit of an extremal black object. In these cases, the weak gravity conjecture suggests that there should be a charged object that the black hole can emit. Such an object corresponds to a particle or brane in the AdS geometry, which remains at rest in the Poincare coordinates if it saturates the bound, and accelerates outward if it strictly satisfies the bound. Since these objects reduce the charge, the resulting solution has a lower vacuum energy (smaller curvature radius) near the horizon, and a larger curvature radius in the UV. Therefore, the charged object demanded by the weak gravity conjecture {\it is} precisely a domain wall satisfying our conjecture. So for any AdS spacetime that can be realized as the near-horizon geometry of a black object, our conjecture follows from a version of the ``usual" weak gravity conjecture.

\section{Argument from eternal inflation}
In attempts to make sense of the dynamics of eternal inflation, a simple criterion emerged:  that the dynamics is not dominated by so-called ``Boltzmann brains," observers who arise spontaneously from vacuum fluctuations in de Sitter spacetime. The basic issue goes back to the paper of Dyson, Kleban, and Susskind \cite{Dyson:2002pf}, who argued that the statistical predictions of eternal de Sitter spacetime are in contradiction to observation (updating an argument originally due to Boltzmann, that the observable universe cannot be a spontaneous fluctuation out of thermal equilibrium).

If we focus on one horizon volume of our vacuum, it contains some ordinary observers like ourselves. If the vacuum survives for a long time, it will also produce Boltzmann brains. If our vacuum lives for longer than about $\exp(10^{92}) $ Hubble times, then ``Boltzmann Earths" will far outnumber ordinary earths \cite{Freivogel:2011eg}. This time scale is ridiculously long, but far shorter than the Poincar\'e recurrence time $\exp(S) \sim \exp(10^{123})$, its lifetime if our vacuum violates our proposed bound.
In this case, a single causal patch of our vacuum would contain super-exponentially more Boltzmann Earths than ordinary Earths,  Boltzmann local groups than ordinary local groups, etc. This is a very serious problem, because in order to do science we assume that we are typical observers.  For example, we assume that we see a typical sample of the statistical fluctuations in the CMB when comparing observation to theory.  This  would not be the case were we to live in  a Boltzmann universe---instead, our observations would be dominated by statistical anomalies and ``miracles'' \cite{Dyson:2002pf}.

A more general analysis of the dynamics of eternal inflation \cite{Bousso:2008hz, DeSimone:2008if} shows that in order for ordinary observers not to be vastly outnumbered by Boltzmann brains, {\it every} vacuum must decay faster than the timescale for it to produce Boltzmann brains. Some vacua may not support Boltzmann brains due to having a very large cosmological constant or exotic (or boring) chemistry. However, vacua that do allow for Boltzmann brains will nucleate them on a timescale much shorter than the recurrence time, since nucleating the brain leads to a small change in the horizon size.  

   Therefore, these vacua must decay on a timescale much shorter than the recurrence time. As a result, the decay must occur via a process that is small compared to the horizon size. Since we expect that the low-energy properties determining whether Boltzmann brains can form are independent from the high-energy processes that determine the tunneling rate, it is natural to conjecture that {\it all} de Sitter vacua with large horizon size decay by such a ``non-gravitational" process.  Note that this condition, while necessary to avoid Boltzmann brains (in vacua that could otherwise produce them), is not in itself sufficient.

Since vacua like ours have an accidentally small cosmological constant, it is presumably a matter of chance that our cosmological constant is positive rather than negative. The stability considerations should be the same for vacua with negative and positive cosmological constant, as long as the cosmological constant is accidentally close to zero. Therefore, {\it all} vacua with horizon size much larger than any other length scale should decay by a process that is small compared to the horizon size. This then implies that AdS vacua with small cosmological constant cannot be completely stable.

\section{Implications for CFTs}
Any completely stable AdS spacetime is believed to be dual to a conformal field theory (CFT). An AdS spacetime that is pertubatively stable but decays nonperturbatively to a lower vacuum is {\it not} dual to a healthy CFT; since the nonperturbative decay can occur anywhere in the spacetime and modifies the boundary after finite time, it corresponds to a pathology in the UV of the would-be dual CFT \cite{Horowitz:2007pr}.  On the other hand, only special CFT's are believed to be dual to AdS spacetimes with semi-classical gravity duals.  The CFT should have large $N$ expansion, with a small number of primary fields with dimensions less than a gap that scales to infinity with $N$ \cite{Heemskerk:2009pn}.

Since we conjecture that AdS spacetimes are at most marginally stable, and that they are unstable if not supersymmetric, this implies that only supersymmetric CFT's can have weakly coupled gravity duals. This is, to our knowledge, true in known examples.  This may merely be due to  our poor knowledge of nonsupersymmetric CFT's, rather than a fundamental limitation.  If so, our conjecture is wrong.
 
The conjecture that all supersymmetric AdS spacetimes are marginally stable also has a consequence for CFTs.  A flat domain wall in the bulk is dual to an RG flow in the CFT.  Therefore, our conjecture implies that all supersymmetric CFTs with gravity duals have at least one supersymmetry-preserving relevant deformation that produces a flow to another supersymmetric CFT.  It would be very interesting to investigate the consequences of this further.

\section{Implications for the Landscape} \label{lands}
Our conjecture, if true, has important consequences for the landscape of string theory vacua. 
Start with two AdS vacua that are connected by a marginally stable domain wall. An arbitrarily large spacetime volume of the ``true" vacuum can be embedded inside a spacetime that is asymptotically the false vacuum.

To be more precise: imagine starting with asymptotically AdS spacetime in the ``false" vacuum with vacuum energy $V_F$. We want to know whether it is possible to build an arbitrarily large region of true vacuum, with vacuum energy $V_T$, while keeping the energy finite. Note that if the false vacuum is unstable to decay, it means that at zero asymptotic energy, a bubble of true vacuum can form inside the false vacuum, and expand to eat up the entire spacetime. Therefore, there exist solutions (with zero energy) that contain arbitrarily large regions of the true vacuum inside the false vacuum. 

When the tension takes on the critical value, it is no longer possible to build a zero energy solution containing the true vacuum. The simple solutions that exist consist of flat domain walls (rather than spherical domain walls) which connect the two vacua. Physically, we may want to ask what this means for building large regions of the true vacuum.

We can address this in the thin wall approximation: we want to build a spherically symmetric domain wall, with the true vacuum inside and the false vacuum outside. The metric inside will be pure AdS with the true cosmological constant; outside will be an AdS-Schwarzschild metric with the false vacuum value of the cosmological constant. 

Solving the Einstein equations across the domain wall leads to the Israel junction conditions, which relate the difference in the extrinsic curvature across the domain wall to the tension of the wall. It is convenient to use the Schwarzschild-like form for the metric inside and outside the domain wall,
\be
ds^2 = -f(r) dt^2  + {dr^2 \over f(r)} + r^2 d\Omega^2~.
\ee
The metric function $f(r)$ takes the form
\be
f_T(r) = 1 + r^2/\ell_T^2 \ \ \ \ \ \ \ \ \ f_F(r) = 1 - M/r^p + r^2/\ell_F^2
\ee
where $p$ is related to the spacetime dimension by $p = D - 3$, and we have absorbed a dimension-dependent order one factor into $M$.

Considering a domain wall that is instantaneously at rest, the junction condition becomes
\be
\sqrt{f_T(r)} - \sqrt{f_F(r)} = {1 \over 2} T r
\ee
Plugging in the formula for the metric functions, moving the false vacuum function to the right side, and squaring yields
\be
1 + r^2/\ell_T^2 + {1 \over 4} T^2 r^2  -  T r \sqrt{1 + r^2/\ell_T^2} = 1 + r^2/\ell_F^2 - M/r^p
\ee
We are interested in how the mass of this object depends on its size. Solving for $M$ gives
\be
M = r^p\left( r^2/\ell_F^2 -  r^2/\ell_T^2 - {1 \over 4} T^2 r^2 +  T r \sqrt{1 + r^2/\ell_T^2} \right)
\ee
Expanding for large $r$ gives
\be \label{larger}
M = r^{p+2}\left( 1/\ell_F^2 - 1/\ell_T^2 -{1 \over 4} T^2 + T/\ell_T \right) + {1 \over 2} r^p T \ell_T + \mathcal{O}(r^{p-4})
\ee
The leading term is negative when the tension satisfies \eqref{conj}, $T/2 < 1/\ell_T - 1/\ell_F$, indicating that a critical bubble with $M=0$ can form and then expand indefinitely to arbitrarily large size.  

In the critical case where the tension saturates \eqref{conj}, the first term in \eqref{larger} vanishes, leaving the subleading term:
\be
M \approx {1 \over 2} r^p T \ell_T = r^p\left(1 - {\ell_T \over \ell_F} \right)
\ee 
The mass per unit area is given by dividing by $r^{p+1}$, so the mass per unit area approaches zero for large bubbles. This is the sense in which a large region of true vacuum can be built inside the false vacuum in the marginally stable case.  Alternatively, rather than constructing solutions with spherical regions of true vacuum, in the critical case one can construct a solution with planar symmetry that interpolates from the lower energy vacuum in the infrared to the higher energy vacuum in the ultraviolet.

Our conjecture suggests that the landscape of AdS vacua is more connected than one might have thought, since starting from any vacuum one can build an arbitrarily large region of another vacuum inside it. One could go further, and speculate that  all AdS vacua can be accessed within a single spacetime; for example, that arbitrarily large regions of every AdS vacuum can be contained inside a single asymptotically Minkowski spacetime. We are not yet prepared to conjecture that this is the case.

For de Sitter vacua,  our conjecture is less surprising, since it is known that transitions between a given de Sitter vacuum and any other location in the landscape are allowed \cite{Brown:2011ry}.  What is novel on the de Sitter side is the conjecture that  vacua with horizon size much larger than any other length scale decay by a process that takes place on distance scales much smaller than the horizon size. Equivalently, the decay of these vacua persists in the limit that the cosmological constant is taken to zero. In the language of  \cite{Aguirre:2006ap}, all de Sitter vacua lie below the ``great divide."

\section*{Acknowledgments}

B.F. was inspired by a talk of Fabrizio Rompineve (based on  \cite{Hebecker:2015zss}) at the ``Beyond the Standard Model" workshop in Bad Honnef, and an ensuing discussion with Enrico Pajer. It is a pleasure to thank in addition Adam Brown, Gia Dvali, Arthur Hebecker, Simeon Hellerman, Albion Lawrence, Matthew Lippert, Massimo Porrati,  and Lenny Susskind for discussions.  The work of MK is supported in part by the NSF through grant PHY-1214302, and he
acknowledges membership at the NYU-ECNU Joint Physics Research Institute in
Shanghai.


\appendix

\bibliographystyle{klebphys2}
\bibliography{ref}

\end{document}